# Automatic cough detection based on airflow signals for portable spirometry system

Mateusz Soliński[1], Michał Łepek[1], Łukasz Kołtowski[2]

[1] Faculty of Physics, Warsaw University of Technology, Koszykowa 75, 00-662, Warsaw, Poland
[2] 1st Department of Cardiology, Medical University of Warsaw, Żwirki i Wigury 61, 02-091, Warsaw, Poland

✉ mat.solinski@gmail.com (M.S.), lepek@if.pw.edu.pl (M.Ł.)

**Abstract.** We give a short introduction to cough detection efforts that were undertaken during the last decade and we describe the solution for automatic cough detection developed for the AioCare portable spirometry system. In contrast to more popular analysis of sound and audio recordings, we fully based our approach on airflow signals only. As the system is intended to be used in a large variety of environments and different patients, we trained and validated the algorithm using AioCare-collected data and the large database of spirometry curves from the NHANES database by the American National Center for Health Statistics. We trained different classifiers, such as logistic regression, feed-forward artificial neural network, support vector machine, and random forest to choose the one with the best performance. The ANN solution was selected as the final classifier. The classification results on the test set (AioCare data) are: 0.86 (sensitivity), 0.91 (specificity), 0.91 (accuracy) and 0.88 (F1 score). The classification methodology developed in this study is robust for detecting cough events during spirometry measurements. As far as we know, the solution presented in this work is the first fully reproducible description of the automatic cough detection algorithm based totally on airflow signals and the first cough detection implemented in a commercial spirometry system that is to be published.

## 1. Introduction

A cough can be described as a sudden, and often repetitively occurring, air expulsion with a forceful expiratory effort. The cough reflex is initiated by irritation of cough receptors in the airways [1]. As a consequence, nerve impulses from the cough center in the brainstem stimulate the diaphragm, intercostal muscles and larynx to produce the explosive expiration of cough. Cough often reflects respiratory irritation or illness and can also occur as an early symptom of asthma, cystic fibrosis or chronic obstructive pulmonary disease (COPD) [1, 2].

Patients suffering from chronic pulmonary diseases should be regularly monitored to evaluate the progress of disease or treatment. A common method for diagnosing and monitoring of pulmonary functions is spirometry. However, patients with COPD frequently complain of breathlessness and cough, which are usually increased during exacerbations. On the other hand, a spirometry maneuver needs physical effort from the patient and can cause irritation of airways that results in cough during the examination. Spirometry standards require that correct spirometry maneuvers do not contain cough and are free of cough artifacts (especially in the first second after the beginning of the forced exhale) [3].

Recently, much effort has focused on developing pocket, mobile peak flow meters and spirometers with the same or similar functionality to stationary clinical spirometers expanding the accessibility of this type of monitoring (e.g., [4, 5]). These systems are designed to perform spirometry measurements with no supervision of physicians or with the supervision of a physician with limited experience (including general practitioners). Therefore, they need automatic real-time algorithms that can detect the cough accurately and efficiently and warn the user in case of incorrect measurement.

The cough detection issue has been exhaustively explored. The growth of computing power allows the analysis of cough signals in real-time using smartphones or dedicated hardware. Most of the research was related to audio (sound) signals [6-23] or accelerometer recordings [24, 25], which remains in contrast to our contribution. In our solution, we do not analyze the sound but the airflow signal passing through the spirometer; thus, the troublesome influence of environmental noise is largely minimized. The main purpose of developing cough detection and segmentation algorithms described in the literature was monitoring a patient's cough over time and counting cough occurrences [6-17]. Some of the research was dedicated to assessing the degree of pathology for patients suffering from cystic fibrosis [18], to detect cold [19], tuberculosis [20] or COPD [21]. There were several studies on the relevance of different sensors for cough detection (e.g., ECG sensor, thermistor, chest belt, oximeter) [26, 27] but no airflow sensor was investigated. With the constant reduction of the size of electronic equipment, there are attempts to develop a wearable cough detection system [17, 28, 29]. A very recent idea is to make use of smartwatches for ambulatory cough monitoring [30]. The identification of common respiratory disorders using cough sounds becomes a clinical tool [31]. Advanced mathematics is also exploited recently to increase the performance of cough detection algorithms, this is e.g. using octonions (octets of real numbers) [32] or so-called Hu moment invariants from image processing domain [33]. The interesting fact is that cough detection was applied not only for human patients but also for veterinary monitoring of farm animals [34, 35].

Most of the literature is related to cough detection in audio (sound) signals rather than in flow signals. There are different purposes and different methods in detecting cough in audio and flow signals. Cough detection in audio signals is used for monitoring cough in time, counting cough events, diagnosing patients basing on long-term cough observations. It usually uses spectral features to analyze audio recordings. Portable and wearable forms of cough detection always use audio (or accelerometer) signals. In turn, cough detection in flow signals is used mainly in spirometry. Only one single spirometry maneuver is then analyzed and assessed. The purpose of this is to improve the quality of spirometry examinations and helping spirometry patients in self-monitoring (as well as helping inexperienced physicians or medical staff) by automatic maneuver assessment [36]. The signals to be analyzed in this case, are short and (generally) non-periodic. The cough detection in airflow signals for spirometry was undertaken in [37] where the authors tried to automatically detect the most important spirometry user errors. They used heuristic features and decision tree models but the description seems to be too brief as only some of the features are described and model parameters are not mentioned. There is also another interesting work dealing with cough airflow signals [38]. In this case, however, the purpose was not cough detection but analyzing voluntary cough signals and predicting whether the full spirometry parameters of a given patient would be above or below the lower limit of normality.

Although cough detection seems to be examined from many different perspectives, reviewing the literature one realizes the low number of patients that produced the records for the dataset, usually not exceeding a dozen, sometimes up to several dozens of subjects. In some cases, each subject produced several cough samples or the recordings of subjects were divided into numerous segments. Therefore, the numerical results presented by the authors may not be always entirely accurate if rescaled for larger or more diverse sets of patients; as such, they may present limited usefulness and reliability, especially in a broad clinical or commercial application. Algorithms for automatic airflow cough detection are implemented in some stationary spirometers, but, the manufacturers do not disclose data on performance or detection methodologies of their solutions, therefore, no data is available for comparative analysis.

In this work, we describe the solution for automatic cough detection developed for the AioCare spirometry system (HealthUp, Poland) [39]. The system consists of three main elements: portable spirometer (class IIa medical device), the mobile application for smartphone and Internet cloud to store the data. During the measurement, the airflow signal is transmitted from the spirometry device to the mobile application where it is analyzed by an algorithm. Clinically important parameters are presented to the user, e.g., forced vital capacity (FVC), forced expiratory volume in the first second (FEV1), their ratio (FEV1/FVC), peak expiratory flow (PEF), etc. Similarly, if any technical errors occurred during the maneuver they are shown to the user who can determine the correctness of the examination and repeat the measurement if needed. The presence of cough is one of the indicators of incorrectness of the maneuver. As the system is intended to be used in a large variety of environments (clinical and in-home) and by both physicians and patients themselves, the cough detection algorithm we present in this work is developed to be accurate and robust and is tested on a large dataset of spirometry airflow recordings. Attention is also paid to the need for the high specificity of the algorithm to avoid negative consequences of incorrect classification as a cough that could cause the unjustified need of repeating the measurement or user's discouragement.

The organization of the article is as follows. In Section 2 the database used for training and validation of the algorithm is briefly described. Preprocessing methods are outlined in Section 3. Section 4 provides an overview of analytical methods adapted to construct the algorithm. The machine learning algorithms used in this work are presented in Section 5. Their results are shown in Section 6. The discussion and summary are given in Sections 7 and 8.

## 2. Data and preparation

The part of the data for the research was obtained from the National Health and Nutrition Examination Survey (NHANES) database by the American National Center for Health Statistics [40]. It is a free data source containing raw spirometry curve data and additional information about the examinations. Spirometry testing procedures of the NHANES database met the recommendations of the American Thoracic Society. Three subsets of the database covering years 2007-2012 were used [41]. The patients were both males and females from 6 to 79 years old. According to the NHANES documentation, participants eligible for spirometry performed an initial first test spirometry examination. Then, if certain criteria were met, a subset of participants performed a repeat second test spirometry exam after inhaling a β2-adrenergic bronchodilator. Multiple individual spirometry curves were typically obtained during both test spirometry examinations. The dataset contains the raw signals for all of these individual spirometry curves. While the majority of spirometry studies collected in NHANES are of high quality, some spirometry curves may show defects such as extra breaths, a cough, a back extrapolated volume error (BEV error) or a false start to the expiratory maneuver. These curves are divided to 4 subsets (A-D) in the NHANES database where subset A contains the curves of acceptable quality, B – curves with a large time to peak flow or a non-repeatable peak flow, C – curves that had either less than 6 seconds of exhalation or no plateau, and D contains cough and BEV error curves. Thus, the cough containing curves were extracted from the D-labeled examinations by 4 experienced human experts to create the dataset of two classes: ATS-acceptable and other error curves versus cough curves. Examples of ATS-acceptable and cough containing maneuvers are shown in Figure 1.

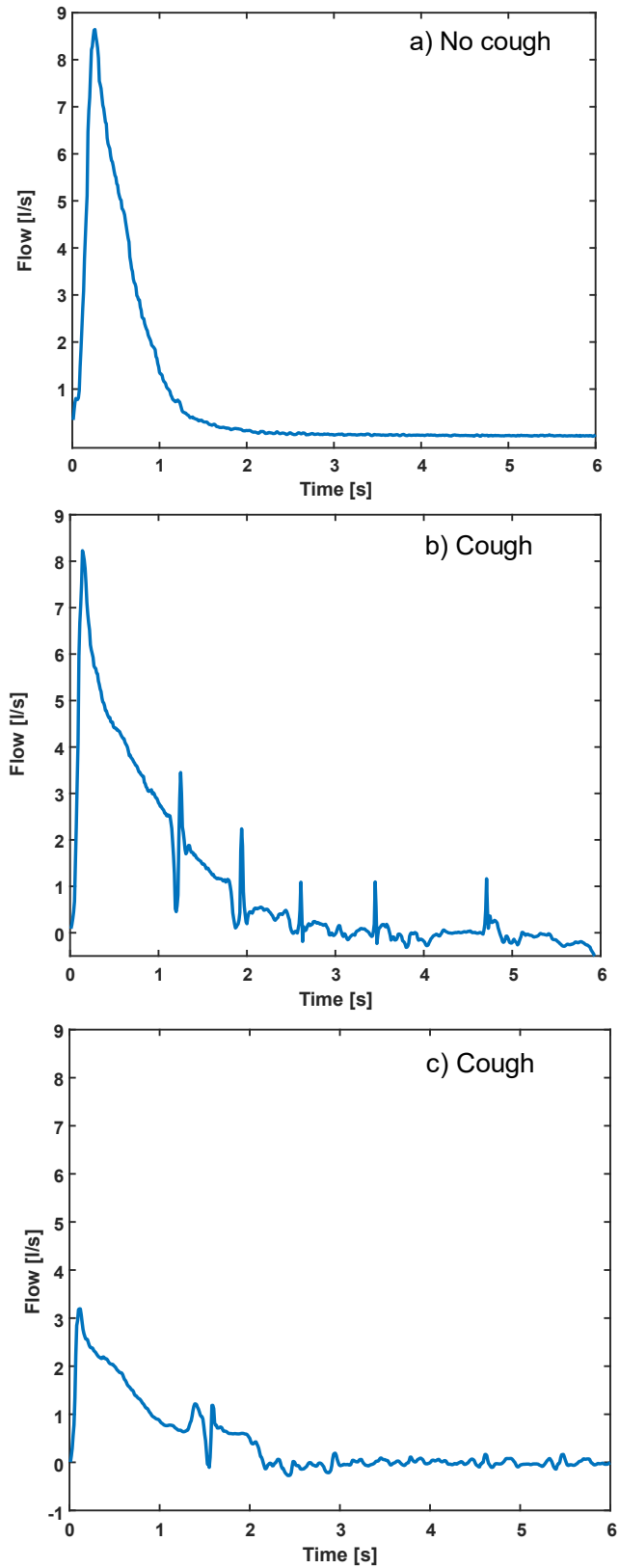

**Figure 1.** Exemplary spirometry maneuvers from the NHANES database: (a) correct ATS-acceptable maneuver; (b) curve containing a very clear cough occurrence; (c) curve containing a less manifested cough occurrence.

Although NHANES data is massive, the cough detection algorithm trained on that data is to run on the signals collected using the AioCare spirometer, hence, the signal collecting devices are different and signal properties may differ (e.g., the sensitivity of airflow sensors, level of noise). Moreover, during the preliminary analysis of the NHANES dataset, it has been found that there are very few non-cough signals with PEF of 1.5 L/s or lower. To prove the reliability of the NHANES-trained and tested algorithm on AioCare-collected curves, the second data set containing 8,939 curves was obtained from AioCare measurements during FVC maneuvers. These signals were obtained from the AioCare database containing the results from the healthy and diagnosed patients suffering from asthma and COPD (with age ranging from 7 to 80) and covered a wide range of PEF values (also <1.5 L/s). To the AioCare additional data set, we have added 19 steady-flow signals generated by Series 1120 Flow Volume Simulator by Hans Rudolph, Inc. Adding curves of this very specific kind to the AioCare data test set was to ensure that the spirometry system will correctly recognize such signals as non-cough ones.

Finally, all data consisted of 19,832 spirometry curves. Table 1 presents an overview of the dataset. It was randomly divided into training (59%), validation (33%) and test (8%) subsets. The sampling rate of all of the data (both NHANES and AioCare) was 100 Hz.

**Table 1.** Data selection extracted from the NHANES database and data obtained from AioCare measurements used in the study.

| Training set | | | |
|---|---|---|---|
| | **NHANES** | **AioCare** | **NHANES + AioCare** |
| **Total** | 8115 | 3604 | 11719 |
| **Cough** | 2570 | 174 | 2744 |
| **Non-cough** | 5545 | 3430 | 8975 |

| Validation set | | | |
|---|---|---|---|
| | **NHANES** | **AioCare** | **NHANES + AioCare** |
| **Total** | 2704 | 3766 | 6470 |
| **Cough** | 884 | 312 | 1196 |
| **Non-cough** | 1820 | 3454 | 5274 |

| Test set | |
|---|---|
| | **AioCare** |
| **Total** | 1643 |
| **Cough** | 105 |
| **Non-cough** | 1538 |

## 3. Data preprocessing

Before extracting features and supplying them to the algorithms, some preprocessing of raw data is needed. These preprocessing steps are to standardize the curves and to clean the region of interest from noise and artifacts. These are performed automatically in the following order:

a. Segmentation of the forced exhale signal from the raw curve. It is usual that the flow curve contains not only the forced exhale but also e.g. inhales before or after the main maneuver. The segmentation covers the fragment from the starting point of forced exhale up to the start of the first inhale (if occurs) that follows the main maneuver.

b. If the length of the forced exhale signal after the segmentation is longer than 600 samples (6 seconds) it is cut down to 600 samples. Due to the spirometry standard [3], a cough that occurs during the first second is the most important as it can significantly change the FEV1 parameter. Cough occurring later than in $6^{th}$ second has very little impact on spirometry parameters (in most examinations the exhaled-volume increase after 6 seconds is residual).

c. Zeroing all of the negative values in the signal. This operation has no effect on extracting features as any of them analyzes the negative part of the signal. However, it lowers the flow-span of the data and zeroes the residual fragments of inhales if they were not entirely extracted during the initial segmentation.

d. Filtering the signal with the moving average of window length of 5 samples (i.e., 0.05 second). Slight filtering is applied to smooth the noise and dispose of minor artifacts.

e. Preprocessing for steady-flow detection. This step recognizes whether the signal is a steady flow signal characteristic for Flow/Volume generator or calibration syringe measurements. The detection is based on testing how many of the samples in the segmented exhale signal differs from the signal median significantly. If this number is low, then it can be assumed that the signal is a steady-flow signal and not a cough (see Table 2 for pseudocode algorithm) and it finishes the classification path for the signal.

**Table 2.** Pseudocode algorithm for determining whether the signal is a steady-flow signal (preprocessing step e). The main idea is to calculate the difference between the signal and its median. If the signal is recognized as a steady-flow signal it is then labeled as non-cough one.

```
signal ← moving mean of signal
for i = 1 to i = length of signal
        if signal(i) < 0.5 * maximal value of signal
        then remove signal(i) from signal
med ← median of signal
for i = 1 to i = length of signal
        signal(i) ← absolute value of (signal(i) – med(i))
sum ← the number of elements in signal where element < 0.1
if sum / length of signal > 0.65
then signal is steady-flow
```

## 4. Feature extraction

Several numerical features have been developed to characterize the presence of cough in a single spirometry curve. The input of the machine learning algorithm consists of 6 features of low computational effort, extracted for each curve. These are:

a. The number of local maxima (spikes) that are longer than 0.05 s and occurred after achieving peak expiratory flow (PEF) (see fig. 2a). The threshold of 0.05 s was determined to separate cough-relevant local peaks from shorter fluctuating ones (noise). Please note that the moving average filtration (applied in Section 3.d) smoothens but usually does not remove peaks (or valleys) if they are clearly visible before.

b. The number of local maxima (spikes) that occurred after achieving PEF with the right-slope amplitude of more than 0.25 l/s (see fig. 2b). This feature counts the peaks that are distinguishable enough from the background and can be markers of cough. The right-slope amplitude is the amplitude between the peak maximum and the first point in time where the first derivative of the signal changes its sign, thus, where the signal starts to increase again.

c. The number of crossings of the signal with horizontal lines (intersections) at 15%, 25%, 50% and 75% of PEF. In this way, 4 separate features are calculated (see fig. 2c), separately for each horizontal line. In non-cough signals, the number of crossings for each horizontal line (if any fluctuations are not present) is equal to 2. This methodology, especially zero-crossing (intersections with x-axis), is widely used in detecting fluctuations in signals from various domains, e.g., in heart rate analysis for both electro- and phonocardiograms [42-45].

The features a–c are graphically outlined in Figure 2.

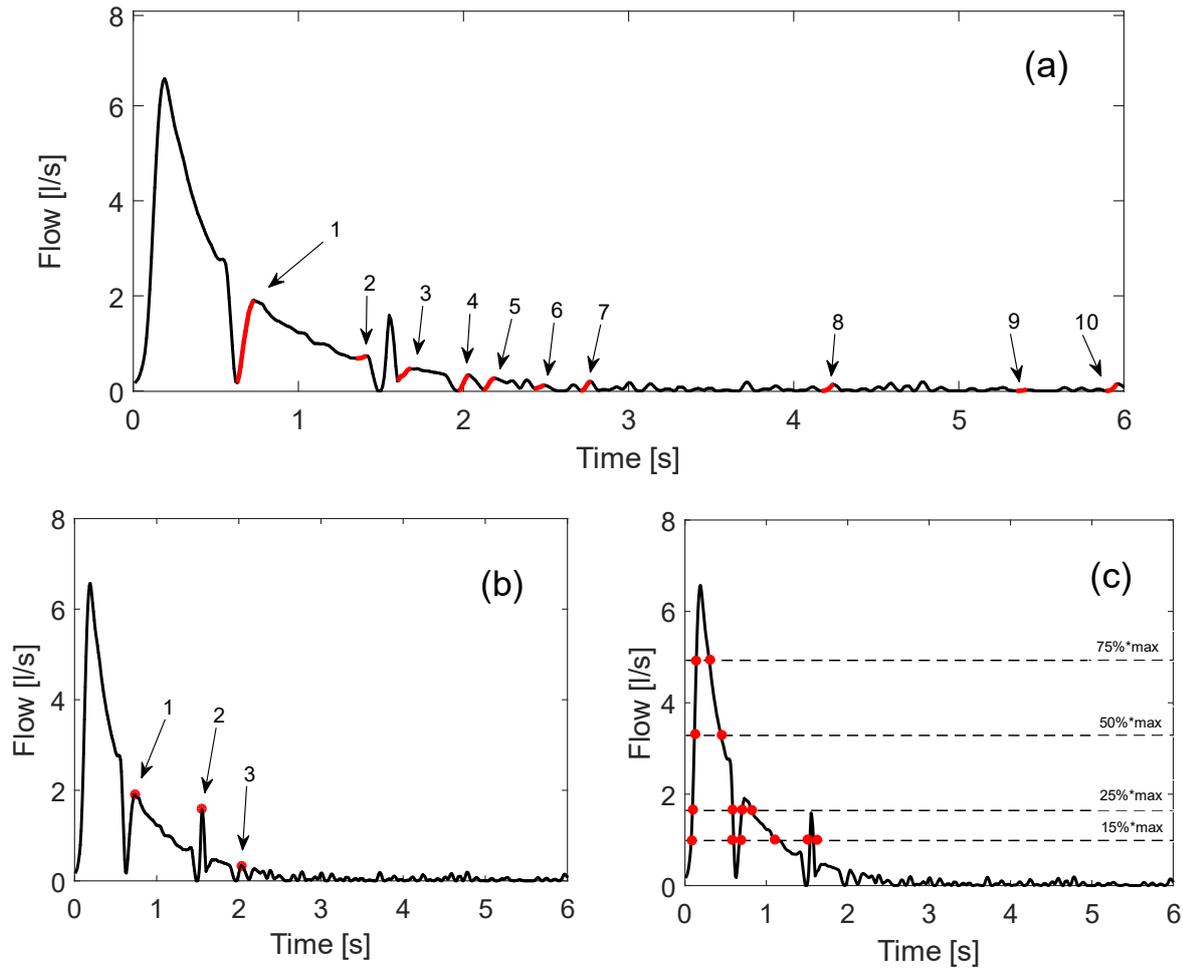

**Figure 2.** Illustration of features derived from the exemplary spirometry curve. (a) Number of local maxima (spikes) that are longer than 0.05 s and occurred after PEF. Subsequent spikes are marked with arrows and denoted with integer numbers. (b) The number of local maxima (spikes) occurred after achieving PEF with the right-slope amplitude of more than 0.25 l/s. There are 3 peaks of interest marked with red dots and arrows. (c) The number of crossings of the signal with horizontal lines (intersections) at 15%, 25%, 50% and 75% of PEF value. The intersections are marked with red dots.

The correlation analysis of features is often useful to determine the relevance and similarity of these features. High correlation (positive or negative) of a specific feature with data labels can indicate high usefulness of this feature in further classification. On the other hand, one should avoid processing features that are highly correlated (close to unity) with each other as it increases the size of the input data and of the model while not providing any additional information. At the beginning of the research we have collected a set of features that could be possibly useful for cough classification (via brainstorming and from the literature). We have discarded the features that were highly correlated with other ones (correlation higher than 0.85). We have performed several training and validation procedures similar to calculating the optimization surface to obtain the final set of features used in the study. A correlation plot for the final features for this study is presented in Figure 3.

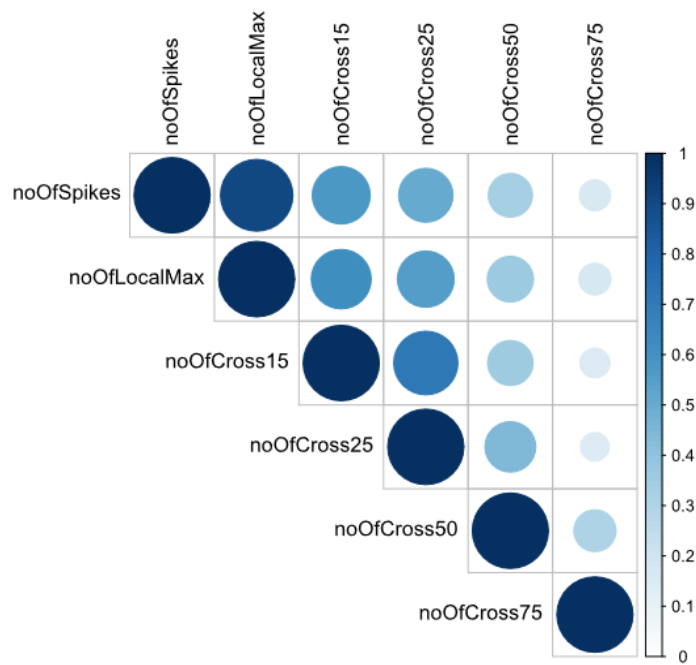

**Figure 3.** Correlation plot of features. Each row and each column show the correlation with other features or cough. The size and color of the circles are proportional to the value of correlation (blue for positive correlation). The number of local maxima with the right-slope amplitude of more than 0.25 L at the descending slope of the forced exhale signal (noOfLocalMax) has significant correlation with the number of spikes that are longer than 0.05 s at the descending slope of the forced exhale signal (noOfSpikes), however, it does not exceed 0.8. The number of crossings of the signal with a horizontal line at 75% of the maximum (noOfCross75) is the feature of the weakest correlation with another ones.

Additionally, each feature was evaluated separately according to its predictive ability using one-way logistic regression models. For each regression model, p-value and odds ratio values were calculated. For all of the features, the p-value is lower than 0.05 which means that these features are statistically significant when used to differentiate between cough and non-cough sets of signals. As statistically significant, the features were analyzed (still separately) with the odds ratio method. The odds ratios were calculated with confidence intervals as follows: 4.09 (95%CI: 3.85-4.34) for the number of spikes, 11.60 (10.52-12.78) for the number of local maxima, 4.06 (3.80-4.35) for the number of crossings at 15%, 5.99 (5.49-6.54) for the number of crossings at 25%, 6.36 (5.76-7.03) for the number of crossings at 50%, and 3.58 (3.30-3.89) for the number of crossings at 75%. For instance, the odds ratio for the number of spikes is 4.09. It means that increasing the number of spikes by 1 would increase the probability of finding a cough in this signal by a factor of 4.09. In this way, we proved (initially) that the selected features would extract information on cough and they could be beneficial if used for building the classifier.

## 5. Machine learning algorithms

All input data (see Table 1) were mean- and standard deviation-normalized before processing to training models. Machine learning models for the study were implemented in the R-Studio environment (version, 1.1.4.5.6, R package version: 3.5.1). Several algorithms were trained and tested to choose one of the highest numerical performance. These were: logistic regression (LR), feed-forward artificial neural network (ANN), support vector machines (SVM) and random forest (RF).

F1 score was used as a metric to select optimal models during optimization and to compare the results. The parameters of the machine learning models were as follows:

a) ANN: Sigmoid function was used as a transfer function in the neuron model. The validation method was k-fold cross-validation (k=5). The number of neurons and the number of iterations were tuned by calculating several combinations of parameters (each combination was run 100 times to obtain reliable statistics). The maximal number of iterations was fit to 60 and 7 hidden neurons were used. The optimization surface is depicted in Figure 4 to visualize the optimization results.

b) LR: LR with k-fold cross-validation (k = 10) was used. The loss function was mean square error.

c) SVM: SVM with a radial-type kernel was used. Model parameters were tuned using grid search among the following parameters: cost of constraints violation (it is the 'C'-constant of the regularization term in the Lagrange formulation) and gamma parameter (kernel parameter). The best parameters were 10 for cost and 0.5 for gamma.

d) RF: We used function based on Breiman's RF algorithm for classification and regression. The number of trees was fit to 400. The number of predictors at each split was established by minimalization of mean square error and were selected as 2. As in the case of ANN, the RF algorithm depends on the initial seed, thus, we have performed 100 runs to obtain reliable statistics.

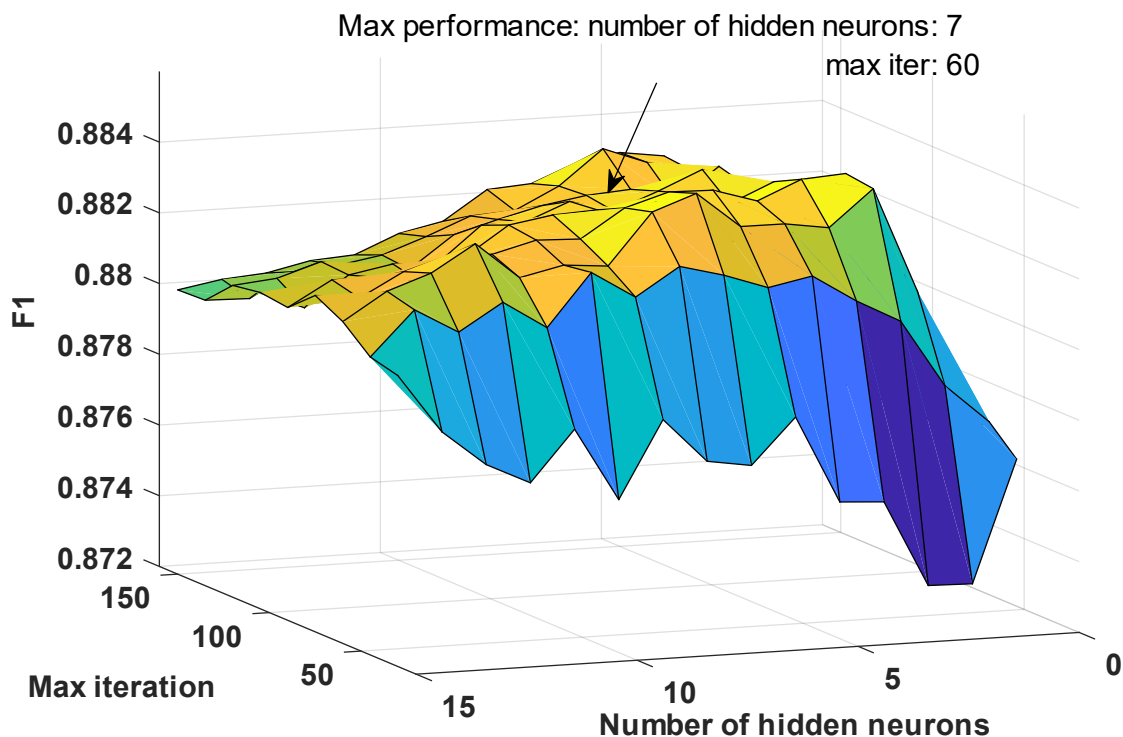

**Figure 4.** The optimization surface for the ANN classifier. The validation procedure was performed using the validation dataset, training time up to 150 iterations and hidden neuron numbers from 1 to 15. The final architecture was 7 hidden neurons and 60 iterations.

The final artificial neural network architecture is presented in Figure 5.

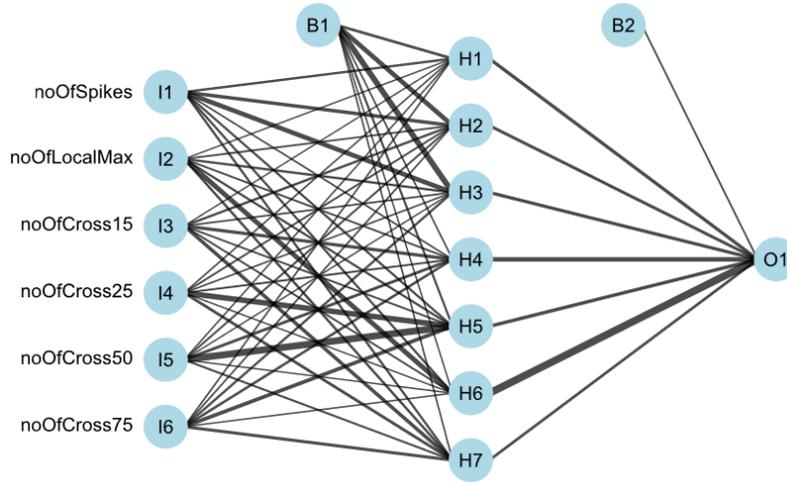

**Figure 5.** Diagram presenting the final ANN architecture with 6 inputs (I1-I6), 7 hidden neurons (H1-H7) and 1 output neuron (O1). Absolute values of weights of inter-neuron connections are proportional to the thickness of the links. B1 and B2 are pictorial representations of the neuron biases.

## 6. Results

Statistical measures used for estimating numerical results of the algorithms were:

a. Sensitivity (or recall) which measures correctly identified actual positives:

$$Sensitivity = \frac{TP}{TP + FN}$$

b. Specificity (or selectivity, precision) which measures correctly identified actual negatives:

$$Specificity = \frac{TN}{TN + FP}$$

c. Accuracy which measures the overall number of correctly classified samples:

$$Accuracy = \frac{TP + TN}{TP + TN + FP + FN}$$

d. F1 score which the harmonic mean of the precision and recall:

$$F1 = \frac{2 \cdot Sensitivity \cdot Specificity}{Sensitivity + Specificity}$$

where $TP$ stands for true positive (correctly identified), $TN$ – true negative (correctly rejected), $FP$ – false positive (incorrectly identified), $FN$ – false negative (incorrectly rejected).

The results for all of the methods are presented in Table 4. For all of the algorithms, the accuracy is roughly similar, however, the other measures vary slightly. ANN achieved the highest scores for sensitivity and F1 score and it was further tested on the AioCare test set to check final performance.

**Table 4.** The results of training and classification on the validation and test sets. Although the results on the validation set are very similar, ANN achieved the highest scores for sensitivity and F1 and it was further tested on the AioCare test set with a satisfactory result. The confidence intervals are also presented in the table.

| | Sensitivity (95%CI) | Specificity (95%CI) | Accuracy (95%CI) | F1 (95%CI) |
|---|---|---|---|---|
| **Validation set (NHANES + AioCare)** | | | | |
| **LR** | 0.756 (0.746-0.766) | 0.969 (0.965-0.973) | 0.929 (0.923-0.935) | 0.849 (0.841-0.858) |
| **ANN** | **0.865 (0.857-0.873)** | **0.941 (0.935-0.947)** | **0.927 (0.921-0.933)** | **0.901 (0.894-0.909)** |
| **SVM** | 0.798 (0.789-0.808) | 0.962 (0.958-0.967) | 0.932 (0.926-0.938) | 0.873 (0.865-0.881) |
| **RF** | 0.795 (0.785-0.805) | 0.968 (0.963-0.972) | 0.936 (0.930-0.942) | 0.873 (0.865-0.881) |
| **Test set (AioCare)** | | | | |
| **ANN** | 0.857 (0.840-0.874) | 0.912 (0.898-0.926) | 0.908 (0.894-0.922) | 0.884 (0.868-0.899) |

As can be seen in Table 4, all of the classifiers we tested, they gave – more or less – similar results of performance. For both validation and test data, the specificity of the algorithm is higher than the sensitivity which can be regarded as a positive property of the algorithm as the minimalization of the number of false-positives was one of the goals in the process of algorithm development.

We also tried to calculate the performance of ANN using balanced training dataset. ROSE package [47, 48] was used for generating synthetic data, i.e., creating new synthetic points from the minority class to increase its cardinality. The balanced training dataset contained 5907 non-cough cases and 5812 cough cases. The best performance obtained for the validation set from grid analysis was achieved for 5 neurons and training time of 30 iterations (F1 = 0.88, Sensitivity = 0.83 and Specificity = 0.92). This performance is slightly worse than in the case of original data due to the decrease in sensitivity result.

## 7. Discussion

We scanned the misclassified samples for different classifiers and found out that, in general, most of the misclassified samples are the same samples. We think that this is understandable because the main impact on the performance should be given by the choice of features, rather than by the choice of classification model. All of the models we used are robust models widely used in machine learning and similar performance may be considered as good and it may mean that none of the models were over-trained. In Figure 6, we present correctly and incorrectly classified samples. There are two cases (false positive and false negative) that were misclassified by all of the models (subfigures e and f). In the first case, it is the spirometry maneuver that has been performed without forced exhalation (low, not smooth, no well-defined maximum), thus, the algorithm misclassified it as a cough as the values of features were high. In the second case, the cough was classified as a non-cough curve. It is probably because most of the curve is smooth and non-cough indeed, and the cough occurs only at the end of the maneuver (two narrow spikes). From a clinical point of view, these two cough spikes at the end of the maneuver would not change FVC significantly, thus, some physicians would probably accept such a maneuver as a usable one.

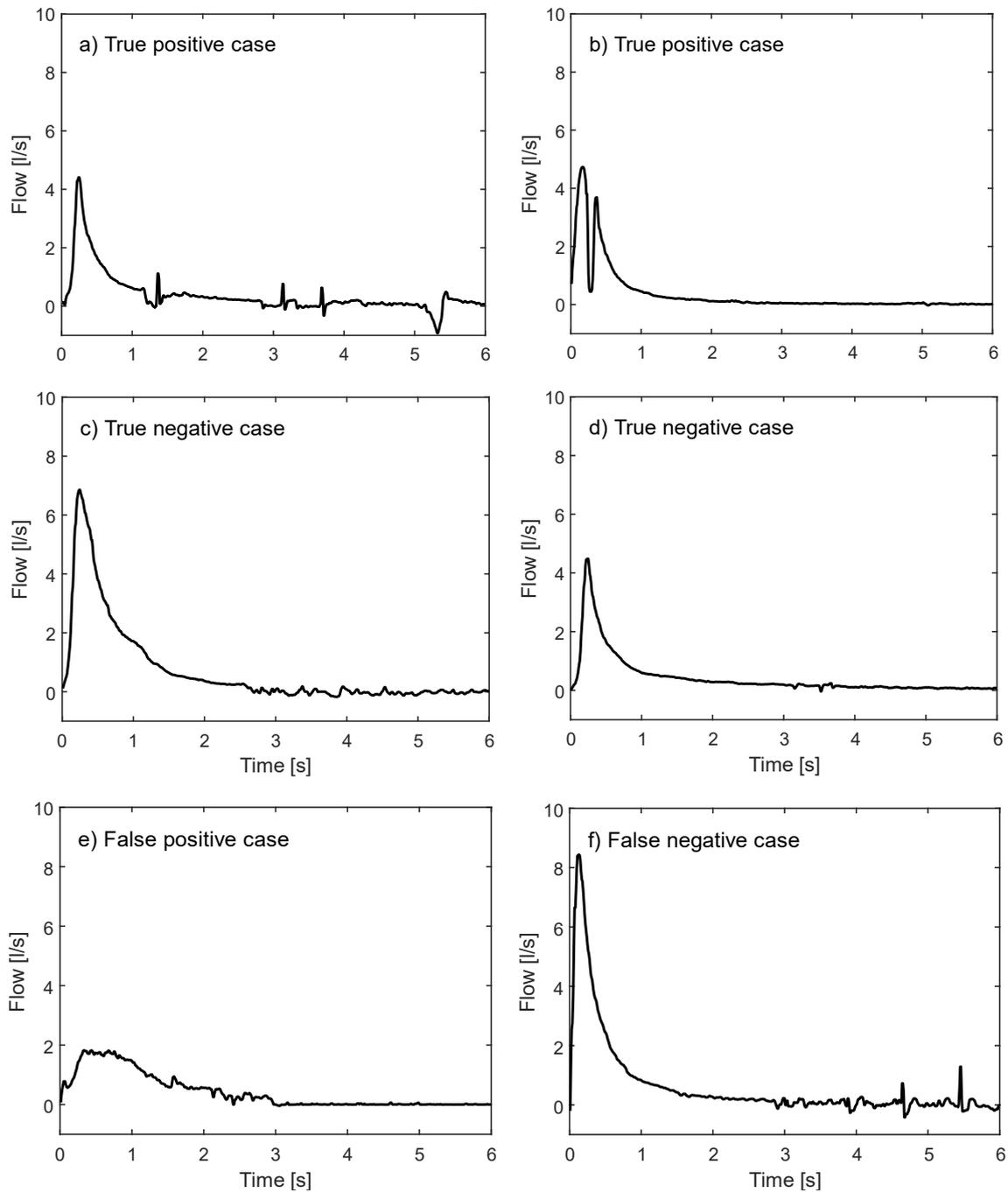

**Figure 6.** Charts presenting correctly (a–d) and incorrectly (e, f) classified curves. In (e), the spirometry maneuver has been performed without forced exhalation (low, not smooth, no well-defined maximum). High values of features caused the algorithm to incorrectly classify it as a cough. In (f), the algorithm misclassified the curve as a non-cough one.

We compared our results to the results obtained in the preceding work [37] where the authors developed airflow-based cough detection as a part of the spirometry user-error detection. The comparison is shown in Table 5. Although the sensitivity of the competing algorithm is higher, we report significantly higher specificity and, thus, higher overall F1 score.

**Table 5.** The final results of the classification algorithm developed in this work in comparison to the results from the competing work [37] dealing with errors in spirometry airflow signals. Although the sensitivity of the competing algorithm is higher, we report significantly higher specificity and, thus, higher overall F1 score.

|  | Our results | Competing work [37] |
|---|---|---|
| **F1 score** | 0.884 | 0.865 |
| **Specificity** | 91.2 % | 81.7 % |
| **Sensitivity** | 85.7 % | 91.9 % |

The resulting specificity of our algorithm is higher than sensitivity, which we found acceptable as the minimalization of false-positive factor was the property of interest due to the functional needs of the application (to prevent situations when the user is informed of a cough error while it did not occur in real). As the ATS standards require, a cough-containing measurement shall be repeated, therefore false positive classifications (low specificity) would force unnecessary maneuvers. We consider this feature as important for a real patient-friendly application because false-positive information would discourage the patient from further maneuvers and examinations.

The performance for the balanced dataset was worse than for the original one. Using the balancing of the data, we lost the information about the appearance frequencies, which is going to affect accuracy metrics themselves, as well as production performance. The cough is relatively rare during spirometry examinations but not extremely rare (i.e., <1% of all cases), thus, we believe that using original data was reasoned. The difference in F1 value between these two approaches (unbalanced, original data versus balanced data using ROSE package) was not very significant; similar performance can suggest that the model is not over-trained. Moreover, the values of the performance parameters obtained for the test data set are only a little lower than the ones for validation set which seems to prove that the model has an acceptable ability to generalize.

## 8. Summary

The detection of cough events in spirometry curves using airflow signals is a tricky task as the cough can be manifested not only in a very clear way but also through small flow disturbances. On the other hand, lots of disturbances can be caused by other processes than cough. We adopted the NHANES database to make sure that the training data is as large and diverse as possible, however, we also used a large number of AioCare-collected signals. The classification algorithm developed in this study is a robust tool for detecting cough events during spirometry measurements and outperforms the previously described approach. Taking into account the full description and reproducibility of the algorithm, we think that our results can be regarded as a noteworthy step forward in clinical and home monitoring in spirometry. The algorithm we developed was implemented in the AioCare mobile application. There are still some curves that were misclassified by the algorithm, however, most of these maneuvers contain small cough disturbances or disturbances which can be similar to cough. Thus, the features developed in this study can be insufficient to distinguish these subtle cases. Further increase in performance and reliability will be the aim of the next works. Another possible extension could be modifying the algorithm to provide information on the time when cough occurred (e.g., in the first second of exhalation) what can be useful for spirometry quality assessment.

## 9. Conflict of interest

Łukasz Kołtowski is the inventor of AioCare and a shareholder of HealthUp. Mateusz Soliński and Michał Łepek are employed by HealthUp.

## 10. Ethical approval

All procedures performed in studies involving human participants were following the ethical standards of the institutional and national research committee and with the 1964 Helsinki declaration and its later amendments or comparable ethical standards.